# Formation of laser plasma channels in a stationary gas

A. Dunaevsky

*Dept. of Astrophysical Sciences, Princeton University, Princeton, NJ 08540, USA*

A. Goltsov

*Kurchatov Institute, TRINITI, Troitok, Russia*

J. Greenberg, S. Suckewer

*School of Engineering and Applied Science, Department of MAE, Princeton University, Princeton, New Jersey 08544, USA*

E. Valeo, and N. J. Fisch

*Princeton Plasma Physics Laboratory, P.O.Box 451, Princeton, NJ 08543, USA*

The formation of plasma channels with nonuniformity of about ± 3.5% has been demonstrated. The channels had a density of $1.2 \times 10^{19}$ cm$^{-3}$ with a radius of 15 μm and with length ≥ 2.5 mm. The channels were formed by 0.3 J, 100 ps laser pulses in a non-flowing gas, contained in a cylindrical chamber. The laser beam passed through the chamber along its axis via pinholes in the chamber walls. A plasma channel with an electron density on the order of $10^{18}$ - $10^{19}$ cm$^{-3}$ was formed in pure He, N$_2$, Ar, and Xe. A uniform channel forms at proper time delays and in optimal pressure ranges, which depend on the sort of gas. The influence of the interaction of the laser beam with the gas leaking out of the chamber through the pinholes was found insignificant. However, the formation of an ablative plasma on the walls of the pinholes by the wings of the radial profile of the laser beam plays an important role in the plasma channel formation and its uniformity. A low current glow discharge initiated in the chamber slightly improves the uniformity of the plasma channel, while a high current arc discharge leads to the formation of overdense plasma near the front pinhole and further refraction of the laser beam. The obtained results show the feasibility of creating uniform plasma channels in non-flowing gas targets.

52.50.Jm, 52.40.Nk

## Introduction

Compression and amplification of laser pulses in plasma due to stimulated Raman backscattering (SRB) is now a proven method for the generation of ultrashort high power laser beams.[1] With the chirped pulse amplification technique (CPA), power densities on the order of $10^{20}$ W/cm$^2$ in the focused beam are available only with meter-size gratings. SRB in plasma overcomes the material limitations of CPA and allows one to achieve power densities several orders of magnitude higher than previously possible with moderate-scale systems.[2] Recently, several schemes of SRB amplification have been proposed.[2,3,4] While some schemes predict efficient SRB of the ionizing laser beam on the ionization front in initially neutral gas,[4] the majority of approaches are based on the interaction between counterpropagating pump and seed beams in a pre-created plasma channel. The resonant Raman backscattering scheme can be extrapolated to high powers conveniently using optical systems.[5]

Being essentially a resonant three-wave process, SRB amplification is rather sensitive to the plasma frequency $\omega_p$. Detuning of the resonance caused by longitudinal fluctuations of the electron density $\delta n_{ez}$ is due to corresponding fluctuations of the electron plasma wave frequency[6] $\delta\omega_p = \omega_p \delta n_{ez}/2n_{e0}$. Here $\omega_p = (4\pi e^2 n_{e0}/m_e)^{1/2}$ is the plasma frequency and $n_{e0}$ is the electron density in the plasma channel. The frequency of the fluctuations should be less than the bandwidth of the pumping beam, which is about twice the SRB linear growth rate $\Delta\omega \sim 2\gamma = a\,(\omega\omega_p)^{1/2}$. Thus, for linear amplification, the maximal allowable fluctuations of the plasma density is

$$\frac{\delta n_{ez}}{n_e} < a_0 \sqrt{\frac{\omega}{\omega_p}}, \qquad (1)$$

where $a_0 = eE/mc\omega$ is normalized electric field in the beam. For successful experimental conditions, the nonuniformity generally ought to be less than a few percent.[7]

For the nonlinear amplification mode, the limitation on allowable density fluctuations is modified. Solodov et al.[6] showed that for given length $L$ of uniform plasma channel the average fluctuation amplitude should be less than

$$\frac{\delta n_{ez}}{n_e} < \frac{2\omega c}{\omega_p^2 \sqrt{L l_\parallel}}, \qquad (2)$$

where $l_\parallel$ is a longitudinal correlation length. For experimental beam parameters and $l_\parallel \approx$ 0.2 mm, the density nonuniformity has to be $\delta n_{ez}/n_e <$ 3-4 %.[5]

In the first experimental arrangement, plasma with a density of $n_e \sim 10^{20}$ cm$^{-3}$ and a temperature of $T_e \sim$ 20 eV was created in a capillary by KrF laser.[8,9] It was shown, however, that the effective length of the plasma channel in copper and LiF capillaries did not exceed 0.2 mm, presumably due to intense pump absorption via an inverse Bremsstrahlung mechanism and initial nonuniformity of the plasma density in the capillary. At lower plasma densities, amplification with $K = \exp(2\gamma L/c) \sim 100$ has been recently demonstrated with Ar-filled glass capillary.[10] However, apparently the significant density nonuniformity decreased the experimentally achieved amplification by several orders of magnitude when compared to the theoretical prediction.

Alternatively, high density gas jets ionized by Nd:YAG laser can provide reproducible plasma channels of 1-2 mm length with a density of $n_e \sim 10^{19}$ cm$^{-3}$ and a temperature of $T_e \sim$ 5 eV. At this order of plasma density, pump absorption due to inverse Bremsstrahlung becomes less important, and amplification ratios up to $K \sim 10^3$ were obtained.[1,11] The possibility to measure directly the density profile in the jet plasma is an additional advantage of the method. However, turbulence in the flow from a nozzle effects the uniformity of the neutral gas density in the flow and, consequently, leads to plasma density nonuniformity. At present, a short effective length $L$ is, probably, the main limitation of the RBS amplification.[1]

Attempting to overcome the difficulties in creating a uniform plasma channel in flowing gas, the ionization of stationary, non-flowing gas can be considered. The creation of tubular plasma channels in stationary, non-flowing gas was first proposed by Durfee and Milchberg for guiding of high intensity laser beams.[12] In a backfill gas at pressures of about 30-200 Torr, plasma channels with an electron density of $1 - 4 \times 10^{18}$ cm$^{-3}$ and a length of ~ 1.2 mm have been formed by Nd:YAG laser beam with a power density of ~ $4 \times 10^{14}$ W/cm$^2$ focused by an aplanatic lens. Longer channels were shown to be possible to create with the use of axicon for linear focusing of the ionizing laser beam. Initially formed with a radial density profile close to uniform, the channel forms a tubular plasma waveguide due to shock expansion, which is rather effective for guiding high intensity laser beams over several Rayleigh lengths.[13] Parameters of such plasma channels were studied experimentally and numerically.[14, 15] Gaul *et al.*[16] reported the creation of 1.5 cm long channel in 400 Torr He backfill by axicon-focused Nd:YAG beam, using either long (~400 ps) 0.6 J pulses or 100 ps, 0.3 J pulses together with a pulsed arc discharge for pre-ionization.

While ionization of stationary gas in a backfill chamber is a rather promising technique for the creation of elongated uniform plasma channels, it is of little use for applications like SRB amplification. High power pump and seed beams, as well as ultra-high power amplified beam, cannot be transported to the plasma channel in dense neutral gas without intense interaction with the gas. For such applications, the gas should be localized in a container placed in an evacuated chamber. The container should have windows to let the interacting and ionizing beams to penetrate into the gas volume.

In this paper, we demonstrate the possibility to create plasma channels several millimeters long, with acceptable uniformity, in a cylindrical gas container having pinholes to let powerful laser beams directly to the plasma channel.

**Experimental setup**

Fig. 1 presents a sketch of the experimental setup. Taking technical details out of consideration, the gas target can be represented as a cylinder with a diameter of 10 mm and a length of 3 mm. The cylinder has two pinholes, P1 and P2 (see Fig. 1), on its end walls to let the laser beams go through the cylinder. The pinholes are made of copper and aligned with the cylinder axis. The optimal diameter of the pinholes is determined by the balance between the minimal possible gas leak out of the cylinder and minimal losses of the ionizing beam due to pinhole closure. Gas is supplied into the cylinder through a needle valve V. The feeding gas pipe has a distributor made of porous metal to decrease possible density nonuniformities induced by the incoming gas flow. Gas pressure in the cylinder is monitored by a miniature pressure transducer. Optical access to the plasma is provided trough a pair of optically flat windows made in the sidewalls of the cylinder. The cylinder is placed in a vacuum chamber with a diameter of about 35 cm on a 5-

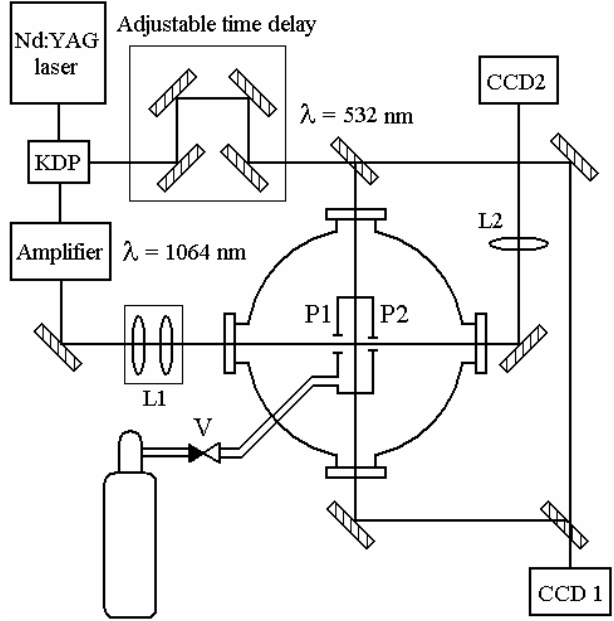

**Figure 1: Experimental Setup**

degrees-of-freedom kinematic stage. The background pressure in the chamber was kept at 300-600 mTorr by a mechanical pump.

An Nd:YAG laser was used to produce an ionizing beam ($\lambda = 1064$ nm) with energy in the range 0.1-0.7 J and a duration of $\tau_b = 100$ ps. An initially rectangular beam of 5 × 1 cm was focused onto the gas target by a pair of cylindrical lenses L1. The beam size in the middle of the gas cylinder was ~ 70 × 70 μm full width at the half maximum (FWHM), which, for estimations, may be represented as a circular beam with a radius of $r_b$ ~ 85 μm. The Rayleigh length, therefore, can be estimated as $Z_R$ ~ 3.8 mm, which is about of the length of the gas cylinder. Thus, we may assume uniformity of power density of the ionizing beam along the channel. Alignment of the ionizing beam with the pinholes was monitored by a camera CCD2 though a telescope L2.

The plasma density distribution in the channel was measured by an interferometer set up on the second harmonic ($\lambda = 532$ nm) probe beam, which was split out of the ionizing beam. Interferograms were registered by a camera CCD1. The radial spread of the plasma was observed by adding a time delay to the probe beam. Phase shift profiles retrieved from interferograms were then processed to restore the radial density distribution by an inverse Abel transform algorithm.[17]

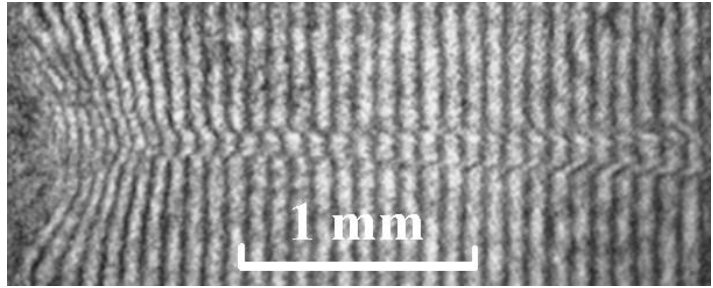

**Figure 2:** Typical interferograms of the plasma channel. Arrow shows the beam direction. $J_{in}$ = 0.38 J, $p$ = 90 Torr $N_2$.

## Experimental results and discussion

### General appearance

A typical interferogram of the plasma channel, taken at a delay of $\tau$ = 800 ps from the arrival of the beginning of the ionizing beam, is presented in Fig. 2. Channels with similar structures were observed in nitrogen in the pressure range of 20 – 300 Torr, in He at the pressures higher than ~ 200 Torr, and in Xe at 30-80 Torr. Channels are formed by laser beams with energies of ~0.2 J and higher. Similar to the backfill and gas jets, an initial plasma spark is formed mainly due to tunneling ionization.[18] Plasma electrons are heated in the laser field due to collisional absorption of laser energy, and drive further ionization by ion-electron collisions. Highly nonequilibrium plasma expands into the neutral gas forming a shock wave and minimum of plasma density at the axis.[10]

While the scenario of the channel formation is similar to backfill and gas jets, several distinctive features have to be mentioned here. First, at the entrance (right side on the interferogram), plasma channels are always wider and shrink gradually along the channel. Second, a bell-like region of dense plasma is formed at the inner surface of the rear wall (left side on the interferogram). Both effects are induced by the presence of the solid walls, where ablative plasma serves as seeding source of hot electrons, and stimulates higher ionization.

**Pinhole size**

The parameters of the ablative plasma on the pinholes determine the limitations of the pinhole diameter. Indeed, the ionizing beam has to penetrate through the pinholes with minimal interaction, thereby presupposing a larger pinhole diameter. Expansion of hot ablative plasma with an overcritical density leads to closure of the pinhole aperture due to beam refraction and reflection, similar to the closure effect in spatial filters. Minimization of pressure disturbances inside the gas container, oppositely, requires smaller diameters. The optimal size should balance these opposite requirements.

The losses on the pinholes are rather significant and reach ~ 40-50 % at the incident energies of $J_{in} \approx 0.5$ J. These losses are caused by closure of the pinhole aperture due to expansion of overcritical ablative plasma formed at the edges of the pinholes. In order to find optimal pinhole sizes, the shock velocity should be estimated. Such an estimation can be done by measuring the energy transmission through the pinhole for different incident beam energies $J_{in}$. The measured values are shown in Fig. 3a by dots for two pinhole diameters: 350 μm and 200 μm. Assuming Gaussian radial and temporal profiles of the beam, energy transmission through the pinholes can be found by integration over $t$ and $r$, taking the pinhole radius decreasing in time with the shock velocity $c_T$:

$$\frac{J_{tr}}{J_{in}}(c_T) = \int_0^{\frac{R_0}{c_T}} \int_0^{R_0 - c_T t} \frac{r}{\tau_b r_b^2 \sqrt{2\pi}} \exp\left(-\frac{t^2}{2\tau_b^2}\right) \exp\left(-\frac{r^2}{2r_b^2}\right) dr dt, \qquad (3)$$

The calculated energy transmission is shown in Fig. 3a by the solid line, from which the shock velocities were estimated. The dependence of $c_T$ vs. $J_{in}$ is shown in Fig. 3b.

Based on the energy transmission, the optimal diameter of the pinholes was picked. For the actual range of beam power density, it corresponded to the diameter of ~ 350 μm for the front pinhole and ~150 μm for the rear one. At smaller diameters of the front pinhole, pinhole losses became too significant, which effects density and uniformity of the plasma channel. A smaller rear pinhole leads to an increase of the density and size of the plasma bubble at the rear wall.

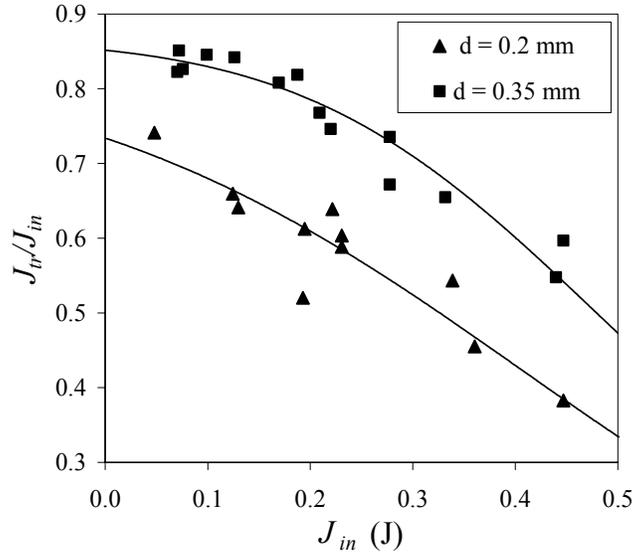

(a)

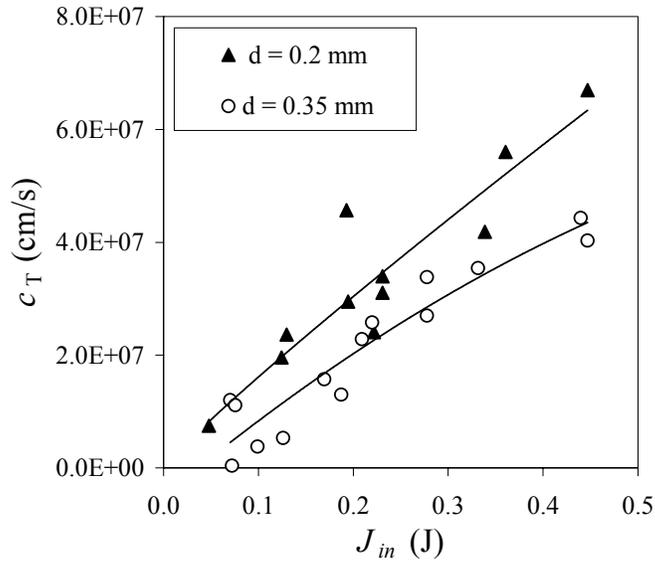

(b)

**Figure 3:** Energy transmission through pinholes $J_{tr}/J_{in}$ (a), and shock velocity $c_T$ (b), versus incident beam energy $J_{in}$. Front and rear pinhole diameters are 350 μm and 200 μm, respectively. Markers in (a) represent measured values, solid lines – calculation via Equation (2).

**Radial profile**

The expansion of the plasma channel in this experiment was found to be similar to the expansion in backfill experiments.[10,11] Indeed, the parameters of the ionizing beam used in the present set of experiments are close to parameters reported by Clark *et al.*[11] and Durfee *et al.*,[10] and non-flowing gas was also ionized. The expansion of the channel in time was studied thoroughly by these authors; thus, we mention here the rather close similarity of our results to the previously reported temporal behavior. Here, the longitudinal changes in the radial profiles of plasma density are important. A typical example of the radial density profiles in 40 Torr of $N_2$ is shown in Fig. 4a for the delay of ~800 ps.

Along the channel, the radial profile changes due to spatially dependent energy deposition. The nonuniformity of the energy deposition may be roughly estimated from the thermal energy $E_{th}$ of the expanding plasma. Clark *et al.*[11] showed the applicability of the self-similar model of expansion of cylindrical shock wave for estimation of $E_{th}$ from the position of the shock wave front: $R(t) = \xi_0 (E_{th}/\rho_0)^{1/4} \tau^{1/2}$. Here $\xi_0 \sim 1$ and $\rho_0$ is the gas density. The longitudinal profile of $E_{th}$ for different gas pressures is shown in Fig. 4b. Thermal energy is high near the front pinhole, where the influence of the wall plasma is high. At pressures lower or about 110-120 Torr $N_2$, the energy deposition is uniform between 0.7 and 2.5 mm and stays within 2-5 mJ/cm depending on pressure. At higher pressures, however, the uniformity disappears, with increased deposition near the entrance of the beam. The increased energy deposition near the front pinhole at higher neutral density might be explained by the increasing influence of the expanding ablative plasma. Hot electrons from the ablative plasma, which penetrate into the neutral gas, might result in the increase of absorption rate at the vicinity of the front pinhole.

Supporting evidence for the hot electron effect comes from attempts to stimulate the dense plasma formation by initiation of an electric discharge in the gas cylinder. While the addition of a low-current glow discharge with the plasma density of about $10^9$ cm$^{-3}$ only improves by a little the uniformity of the channel, a high current arc discharge leads to formation of a dense, highly nonuniform plasma bubble at the front pinhole, resulting in its complete closure.

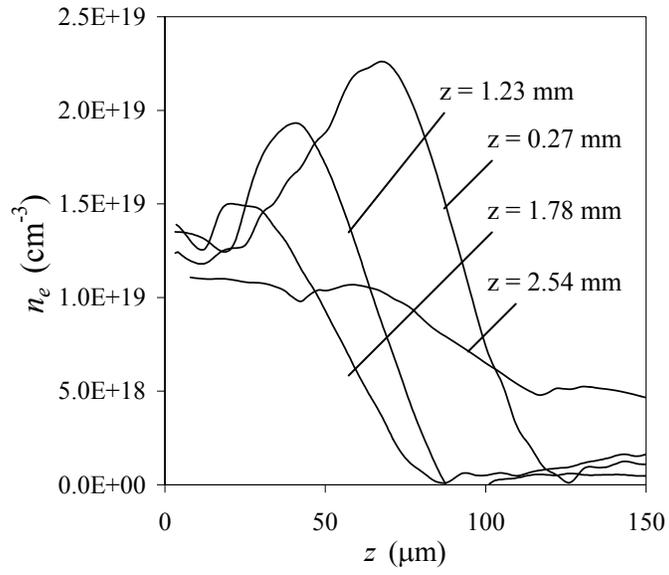

(a)

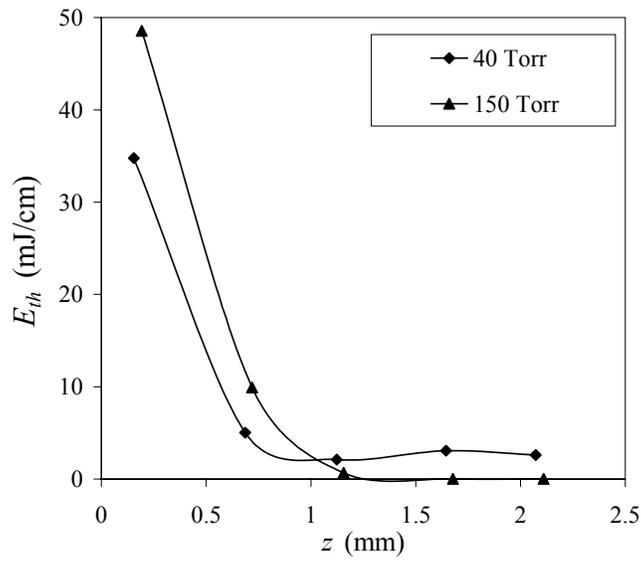

(b)

**Figure 4:** Radial profiles of the plasma density at $p = 40$ Torr $N_2$ (a), and deduced thermal energy $E_{th}$ along the channel (b). $J_{in} = 0.43$ J, $t = 800$ ps.

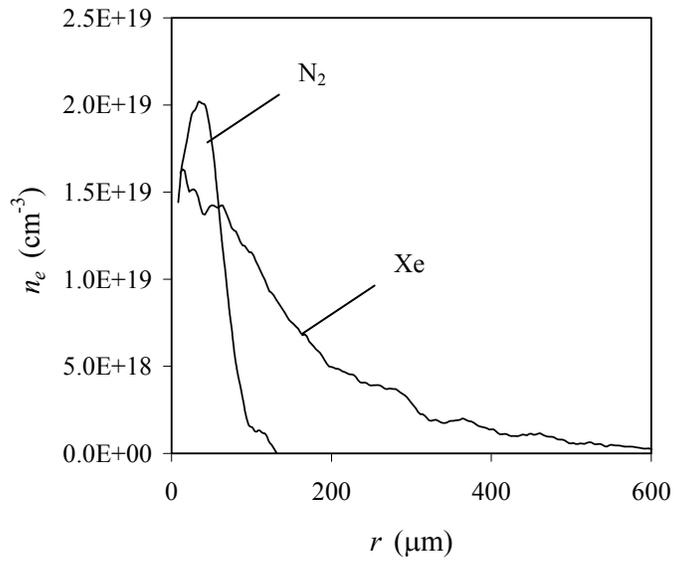

**Figure 5:** Radial profiles of the plasma density in $N_2$ and in Xe at p = 60 Torr. $J_{in}$ = 0.41 J, t = 800 ps, z = 0.4 mm from the front pinhole.

The radial expansion of the shockwave with formation of the minimum plasma density at the axis was observed in nitrogen, helium, and argon. In xenon, however, the radial profiles of the plasma density differ from those lighter gases (see Fig. 5). Even after about 1 ns from the ionizing laser pulse, the profile keeps a Gaussian-like shape. The density of the channels in Xe is about the same as for other gases, but at lower neutral densities. Channels in Xe are usually nonuniform and shorter than the cylinder length. This may be explained by the lower ionization potentials and higher mass of Xe atoms (which affects the expansion and ionization), and by the influence of the ablative plasma from the front pinhole.

Differences between gases are illustrated also by the transmission of an ionizing beam energy through the gas cylinder. For high-Z gases, energy transmission falls to about 10% at about 80-100 Torr and remains at the same level at higher pressures for all gases (see Fig. 6). In helium, however, minimal transmissions of 31-33 % were observed at the pressures higher than 300 Torr. The pressure at which the energy transmission reaches its minimal value coincides well with the maximal pressure at which plasma channels with acceptable uniformity may be created. At higher pressures, the plasma becomes denser near the front pinhole, and the thermal energy $E_{th}$ becomes low and decreases significantly along the channel.

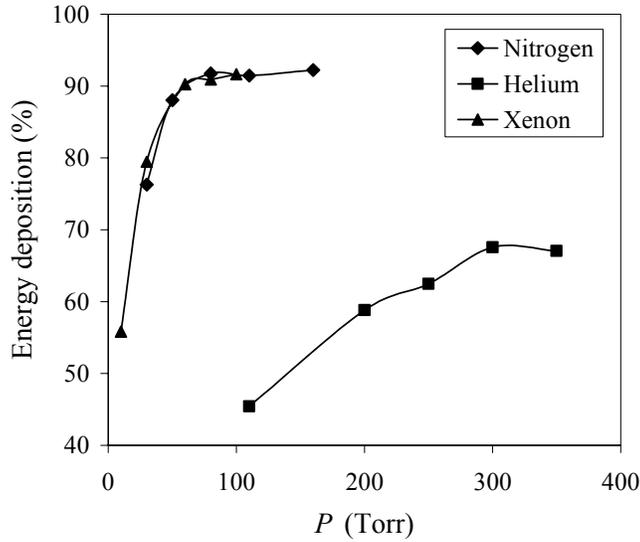

**Figure 6: Total energy deposition for different gases.**

**Longitudinal profile**

The uniformity of the plasma channel is optimized at the axis at a certain time from the channel formation. Indeed, the higher the initial density, the higher the expansion rate and the consequent density drop at the axis. Thus, at some point in time, the initially nonuniform channel will have uniform density along its axis. This effect is particularly prominent for nonuniformities induced by the presence of ablative plasma at the entrance and the exit of the gas cylinder. For instance, we found that in nitrogen the axial part of the channel has good uniformity at about 800 ps from the ionizing pulse. The corresponding longitudinal density profiles are presented in Fig. 7. At $P = 40$ Torr, the channel had an average density of $7\times10^{18}$ cm$^{-3}$ with a nonuniformity of about $\pm 7.5\%$ and a radius of less than 15 μm. The best result was observed at $P = 40$ Torr, when a channel with nonuniformity of about ± 3.5% and had a density of $1.2\times10^{19}$ cm$^{-3}$ with a radius of 15 μm.

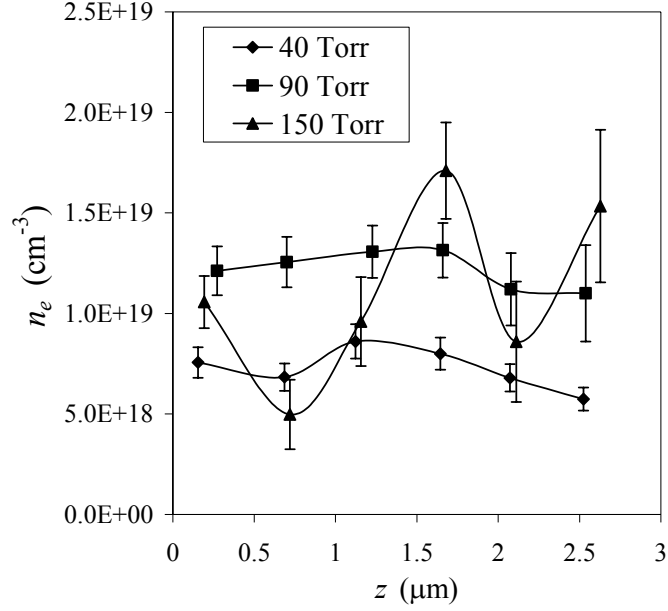

**Figure 7:** Longitudinal profiles of the plasma channel in $N_2$. $J_{in}$ = 0.43 J, t = 800 ps.

The structure of the plasma channel with the average depth of $\delta n_{er} \sim$ 40% (see Fig. 4a) might be attractive, for instance, for SRB schemes with strong plasma wave localization.[19,20] It was shown that in a multimode regime, continuum modes of the plasma channel might contract the backscattered beam and increase the growth rate.[21] For our parameters of the channel (density at the axis $n_{e0} \sim 1.2 \times 10^{19}$ cm$^{-3}$ and the average depth of the channel $\Delta n_e \sim$ 40%), the plasma density gradient exceeds the bandwidth of SRB: $\eta = (\delta n_{er}/ n_{e0})/(2\gamma/\omega_{p0}) \sim$ 20-50 for beams with a power density of about $10^{14}$ W/cm$^2$. If $\eta > 1$, strong plasma wave localization occurs, and SRB amplification would benefit from contraction. Indeed, beams with $\lambda$ = 0.88 nm and $I \sim 5\times10^{14}$ W/cm$^2$ in the plasma channel will fall into appropriate conditions for the continuum mode contribution: $U_1 < a_0 < U_0$.[21] Here $U_0 = (\delta_r n_e/ n_{e0})/2(\omega/\omega_{p0})^{1/2} \sim$ 0.17, $U_1 = U_0/(\pi^{1/2}(2(\omega/\omega_{p0})3^{/4}) \sim$ 0.01, and $a_0 = eE/mc\omega \sim$ 0.02.

## Conclusion

The formation of plasma channels with high uniformity has been demonstrated. The channels were formed by 100 ps Nd:YAG laser pulses with an energy of about 0.3 – 0.5 J in a non-flowing gas, contained in a cylindrical chamber. The laser beam passed through the chamber along its axis via pinholes in the chamber walls. The plasma channel, with an electron density in order of $10^{18}$ - $10^{19}$ cm$^{-3}$, was formed in pure He, $N_2$, Ar, and Xe. A uniform channel forms at proper time delays and in optimal pressure ranges, which can depend on the gas molecular weight. While the influence of the laser beam interaction with the gas leaking through the pinholes was found to be insignificant, the ablative plasma on the walls of pinholes by the wings of radial beam profile is crucial in the propagation of the ionizing beams, and consequently in the channel formation and its uniformity. Low-current glow discharges initiated in the chamber may improve the uniformity of the plasma channel.

## Acknowledgements

Authors would like to acknowlege Dr. A. Morozov for his support and fruitful discussions. Technical assistance of N. Tkach is greatly appreciated. This work was supported by the United States Defense Advanced Research Projects Agency (DARPA).